\documentclass[a4paper, twocolumn, showpacs, nofootinbib, amssymb]{revtex4}

\usepackage[intlimits]{amsmath}

\begin{document}

\title{Light incoherence due to background space fluctuations}

\author{Michael~Maziashvili }
\email{maziashvili@iliauni.edu.ge} \affiliation{School of Natural Sciences and Engineering, Ilia State University,\\ 3/5 Cholokashvili Ave., Tbilisi 0162, Georgia}

\begin{abstract}

Working by analogy, we use the description of light fluctuations due to random collisions of the radiating atoms to figure out why the reduction of the coherence for light propagating a cosmological distance in the fluctuating background space is negligibly small to be observed by the stellar interferometry.

\end{abstract}

\pacs{04.60.Bc }


\maketitle

\subsection{Introductory remarks}

With general relativity  (GR) space-time became a dynamical variable. Then it is natural to assume that the background space undergoes quantum fluctuations that affect the matter fields propagating in it. Taking into account that in view of the basic principles of quantum mechanics (QM) and GR an inevitable uncertainty is expected to attend the measurement of any distance \cite{Osborne:1949zz, Anderson, Wigner:1957ep, Salecker:1957be, Regge:1958wr, Peres:1960zz, Mead:1964zz, Mead:1966zz, Karolyhazy:1966zz, Padmanabhan:1987au, Ng:1993jb, AmelinoCamelia:1994vs, Sasakura:1999xp, AmelinoCamelia:1999zc}, one might suggest that  the wave-length of the propagating wave undergoes fluctuations as well. On the dimensional grounds one could parametrize these fluctuations as: $\delta \lambda = \beta l_P^\alpha \lambda^{1-\alpha}$, where $\beta$ is a numerical factor of order unity, and two particular values of $\alpha$ that have more or less clear physical motivations are: $\alpha = 2/3$ \cite{Karolyhazy:1966zz, Ng:1993jb, Sasakura:1999xp} and $\alpha = 1/2$ \cite{AmelinoCamelia:1999zc}. Based on this fact, an intriguing idea for detecting the space fluctuations was proposed in \cite{Lieu:2003ee}. The idea is to consider a phase incoherence of light coming to us from extragalactic sources. The authors of \cite{Lieu:2003ee} noticed that albeit the frequency fluctuation $\delta\omega = 2\pi \delta \lambda /\lambda^2$ is small, the product $t\omega$ which enters the plane-wave can acquire appreciable shift, $t\delta\omega$, when $t$ is large enough. The approach put forward in \cite{Lieu:2003ee} oversimplifies the actual physical situation in that the authors judge the light coherence basically on the basis of the superposition of two monochromatic plane waves one of which acquires the phase shift, $t\delta\omega$, and conclude that when this phase shift approaches $\pi$ - an interference fringe pattern should be destroyed. Because of this approach one finds an enormous effect - one of the reasons for which is clearly the absence of an actual working model for estimating the degree of coherence. The attempt to merely reduce the phase shift $t\delta\omega$ by saying that the fluctuations must add up more slowly - say as $\sqrt{t\omega}(\delta \omega/\omega)$ (see section C) - does not justify this approach. In view of the above comments, let us notice that the concept of a chaotic light \cite{Loudon} might be useful for qualitative understanding of the actual effects of background space fluctuations on the light beam. For this reason, let us first recall some basic points concerning the coherence of light, which is subjected to the random process of collisions of the radiating atoms.

\subsection{Coherence of the fluctuating light caused by the collisions of the radiating atoms}

Usually, the atoms of the radiating medium undergo the random collisions with each other. On the other hand, the total field coming from this atomic-gas is a superposition of a huge number of fields, {\tt Einzelwellen}, - one for each radiating atom. When the radiating atom undergoes a collision - the phase of the radiated field suffers the random jump. As the collisions have the random nature - the {\tt Einzelwellen} acquire the random phases: $e^{i\theta_j(t)}$. The correlation function at a fixed point (some distance away from the radiating source), where the time-dependence of the field is measured, takes the form \cite{Loudon} 

\begin{eqnarray}\label{Zufallsphase}
\left\langle \Phi^*(t) \Phi(t+\tau) \right\rangle \,=\, N\left\langle \Phi_j^*(t) \Phi_j(t+\tau) \right\rangle  \,= \nonumber \\  N\left\langle \Phi_j^*(t) \Phi_j(t+\tau) \right\rangle_0   \left\langle e^{i[\theta_j(t+\tau)-\theta_j(t)]}  \right\rangle ~,
\end{eqnarray} where one has taken into account that the phase angles of the waves from different atoms have the different random values and the cross terms give a zero average contribution, respectively. The correlation function for the beam as a whole is thus determined by the {\tt Einzelatom} contributions. In Eq.\eqref{Zufallsphase} $N$ stands for the number of atoms, the average is understood in the statistical sense and subscript $0$ indicates the correlation function in absence of the collisions. To estimate the correlation function over the random-phase angles in Eq.\eqref{Zufallsphase}, let us notice that this quantity depends on the collision probability which in general can be written as $W(\tau/\tau_0)$, where $\tau_0$ stands for the mean period of a free flight. That is, $d\tau W(\tau/\tau_0)$ is the probability the atom to move freely in the time interval $\tau, \tau+d\tau$. The phase angle of an {\tt Einzelwelle} jumps to a random value after its source atom suffers a collision and the average in Eq.\eqref{Zufallsphase} subsequently vanishes. So, the average over the random-phase angles in Eq.\eqref{Zufallsphase} depends on the number of collisions and can be estimated immediately by the probability density as

\begin{eqnarray}\label{Wahrscheinlichkeitsdichte}
\left\langle \Phi^*(t) \Phi(t+\tau) \right\rangle \,=\, \left\langle \Phi^*(t) \Phi(t+\tau) \right\rangle_0   \int\limits_{\tau}^\infty d\xi W(\xi/\tau_0)  ~.
\end{eqnarray} This equation implies that the atoms that have experienced collisions make no contribution to the correlation function. It can be readily understood. Namely, recalling that during the collision the {\tt Einzelwelle} can acquire an arbitrary phase within the interval $(-\pi, \pi)$, the average value of the phase under assumption that $N_c(\tau)\equiv N\int\limits_{0}^\tau d\xi W(\xi/\tau_0) \gg 1$ (here $N_c$ denotes the number of atoms that have experienced collisions before the elapse of a time $\tau$) results in   

\begin{eqnarray}\label{Gleitender Mittelwert}
\left\langle \cos\theta \right\rangle \,=\, \frac{1}{N_c} \sum_{j=1}^{N_c}\cos\theta_j \,=\, \nonumber \\ \frac{1}{2\pi} \sum_{j=1}^{N_c}\frac{2\pi}{N_c}\cos\theta_j  \,\simeq\,  \frac{1}{2\pi} \int_{-\pi}^\pi d\theta \,  \cos\theta \,=\, 0  ~. 
\end{eqnarray} We are now in a position to address the question posed in the title of this article.

\subsection{Coherence of the light affected by the metric fluctuations}

Let us now apply the lessons learned from the previous section for estimating the degree of light coherence in presence of the background metric fluctuations. The Eq.\eqref{Wahrscheinlichkeitsdichte} can readily be applied to our problem. As we are dealing with the small fluctuations of the gravitational field, it implies the linearized equations of motion and we can assume that the effect of background space fluctuations are mutually independent for each {\tt Einzelwelle}. The phase fluctuations for {\tt Einzelwellen}, starting from the zero value, are growing with time in a stochastic manner. The meaning of the $\int\limits_{\tau}^\infty d\xi W(\xi/\tau_0)$ function for the problem we are dealing with is as follows. Roughly, for a given instant of time, the phase fluctuations are uniformly occupying some interval\footnote{It would be more natural to assume the Gaussian distribution (having the width $2\delta\omega/\omega$) instead of the uniform one, but for the accuracy of our discussion it is less essential.} $(-\delta\theta(\tau), \delta\theta(\tau))$. The suppression of the correlation function is negligible for $\delta\theta(t) \ll \pi $ and becomes appreciable when $\delta\theta(\tau)$ becomes $\simeq \pi$, see Eq.\eqref{Gleitender Mittelwert}. In view of this judgment, the function $\int\limits_{\tau}^\infty d\xi W(\xi/\tau_0)$ may be understood as the probability that, before the elapse of a time $\tau$, the size of the phase fluctuation (for {\tt Einzelwelle}) will be comparable to $2\pi$. Furthermore,  one can notice that the physics we are dealing with is quite analogous to the Brownian motion \cite{Chandrasekhar:1943ws} and assuming that the probability distribution has the Gaussian form with the mean square value of fluctuation proportional to time - one can write down an explicit form of $\int\limits_{\tau}^\infty d\xi W(\xi/\tau_0)$ and make estimations. Here $\tau_0$ stands for the time-scale at which the root of the mean square fluctuation becomes comparable to $2\pi$. Let us notice that the precise form of $W(\xi/\tau_0)$ is less important for our discussion. One just needs to convince himself that the function $\int\limits_{\tau}^\infty d\xi W(\xi/\tau_0)$ becomes very small as $\tau$ approaches $\tau_0$ and is very close to unity for $\tau \ll \tau_0$. However, for the sake of further simplicity and clarity, we will not follow this way - but rather take a more straightforward approach.

Let us first estimate the scale $\tau_0$ in Eq.\eqref{Wahrscheinlichkeitsdichte}. As it was mentioned above, it denotes the time during of which the root of the mean square value of the phase-shift becomes of the order of $2\pi$. With respect to the discussion put forward in \cite{Lieu:2003ee}, the frequency fluctuation of the order of $\delta\omega\simeq l_P^\alpha\omega^{1+\alpha}$ is understood to take place at each wave-length, that is, sequentially with the time-intervals $\omega^{-1}$. Hence, during the time interval $\tau$, the phase undergoes $n=\tau\omega$ steps of fluctuations. As the fluctuations add up randomly, the phase shift accumulated during the $\tau_0$ will be not $\tau_0\delta\omega\equiv\omega^{-1}\delta\omega n$ but rather \cite{Chandrasekhar:1943ws}    

\begin{equation}\label{Maßstab der Zeit}
 \omega^{-1}\delta\omega\sqrt{\tau_0\omega} \,=\, 2\pi ~, ~~ \Rightarrow ~~  \tau_0   \,=\, \frac{(2\pi)^{2+2\alpha}}{\beta^2 l_P^{2\alpha}\omega^{1+2\alpha}}  ~.
\end{equation} In what follows we will assume that at a time $\tau$ the phases of {\tt Einzelwellen}, $\theta_j$, uniformly occupy the region $(-\delta\theta(\tau), \delta\theta(\tau))$, where $\delta\theta(\tau) = \omega^{-1}\delta\omega\sqrt{\tau\omega} $. Though this assumption might seem more natural for the problem under consideration, it should be remarked that in deriving the Eq.\eqref{Maßstab der Zeit} we have not used precisely this form of probability distribution. This assumption, that the phase undergoes the fluctuations at constant intervals of time $\omega^{-1}$ apart - in such a way as to have the probability at each occasion given by the uniform distribution,

\begin{eqnarray}
\varpi(\theta)= \begin{cases}
    \omega/2\delta\omega       & \quad \text{for } -\delta\omega/\omega \leq \theta \leq \delta\omega/\omega \\
    0  & \quad \text{if } |\theta| > \delta\omega/\omega  \\
  \end{cases} ~, \nonumber 
\end{eqnarray}

\noindent  leads to somewhat more complicated picture. In this case, for the probability $W_n(\theta)d \theta$ that after a time $\tau=n\omega^{-1}$ the phase of the {\tt Einzelwelle} will be found in the interval $(\theta - d\theta/2, \theta+d\theta/2)$, one arrives at the expression \cite{Chandrasekhar:1943ws}  

\begin{eqnarray}
W_n(\theta) \, \propto \, \int_{-\infty}^\infty d\xi  \, \left[\frac{\sin\xi}{\xi} \right]^n \cos\left(\frac{\xi\theta\omega}{\delta\omega}\right) ~. \nonumber
\end{eqnarray} It would be more precise to estimate the mean square fluctuation of phase as a function of time by using this distribution, but, for $\delta\omega/\omega$ is very small - such a rigor and precision is needles for our discussion. Going on, for the suppression factor of the correlation function one finds

\begin{equation}\label{Schwingungsverteilung}
\left\langle \cos\theta \right \rangle \,=\, \frac{1}{2\delta\theta(\tau)}\int_{-\delta\theta(\tau)}^{\delta\theta(\tau)} d\xi \cos\xi \,=\, \frac{\sin\delta\theta(\tau)}{\delta\theta(\tau)} ~. 
\end{equation}

\noindent  From Eq.\eqref{Schwingungsverteilung} one sees that the suppression is appreciable when $\tau$ becomes of the order of $\tau_0$ and is negligibly small for $\tau \ll \tau_0$.

Now let us see what other corrections due to background space fluctuations might be expected and then summarize the overall effect on the light coherence coming from distant (extended) source.

The light signal from astrophysical sources can be described by the (spherical) wave-packet \cite{van Cittert-Zernike}  

\begin{eqnarray}\label{Kugelwelle}  \Phi(t,\, r) \,=\,  \int\limits_{-\infty}^\infty d\omega
\,\Xi(\omega) \, \frac{e^{i\left[k(\omega)\,r-\omega
    t\right] }}{r} ~, \nonumber  \end{eqnarray} where the function $\Xi(\omega)$ is understood to differ appreciably from zero only within a narrow range, $\bar{\omega}-\Delta\omega < \omega < \bar{\omega}+\Delta\omega$, around a mean frequency: $\bar{\omega}$. Here we assume the modified dispersion relation $k(\omega)$ which is understood as a consequence of the background space fluctuations. The frequency fluctuation $\delta\omega \simeq  l_P^\alpha\omega^{1+\alpha}$,  indicates that the amplitude $\Xi(\omega)$ can be known approximately. Taking into account the random nature of fluctuations, one can estimate the average value of the amplitude as 
    
\begin{equation}\label{Neudefinition}
    \mathfrak{A}(\omega)  \,=\, \int\limits_{-\infty}^\infty d\xi \, \frac{\Xi(\xi)}{\sqrt{2\pi}\delta\omega} \, e^{-\frac{(\xi-\omega)^2}{2\delta\omega}} ~.
\end{equation}

It is worth mentioning that in general a small perturbations of the Fourier coefficients may cause a huge cumulative effect that must be properly regularized before using it as a physically meaningful result \cite{Tikhonov-Arsenin}.  As a simple illustrating example, let us consider the function

   \[ f(t) \,=\, \sum\limits_{n=0}^{\infty} a_n \cos(nt) ~, \] and assume that its coefficients are perturbed in the following way: $\widetilde{a}_n = a_n + \varepsilon/n$ for $n \geq 1$ and $a_0=\widetilde{a}_0$. It is plainly seen that the net difference between the coefficients with respect to the metric $l_2$ 
   
   \[ \left[\sum\limits_{n=0}^{\infty} \left(\widetilde{a}_n \,-\, a_n \right)^2  \right]^{1/2} \,=\, \varepsilon  \left[\sum\limits_{n=1}^{\infty} \frac{1}{n^2}  \right]^{1/2} \,=\, \varepsilon \sqrt{\pi^2/6} ~, \] can be made arbitrarily small at the expense of $\varepsilon$. On the other hand, the deviation of the perturbed function
   
   \[ \widetilde{f}(t) \,-\, f(t) \,=\, \varepsilon\sum\limits_{n=1}^{\infty} \frac{\cos(nt)}{n} ~,  \] may be arbitrarily large as for $t=0$ this series merely diverges. So, in practice for reading the field through the Fourier modes this effect should not be taken as a physical one but rather one needs some (optimal) method of regularization \cite{Tikhonov-Arsenin}. We will not face this sort of problem in what follows - but nevertheless it is instructive as it can become the source of the similar misconception entailed by the paper \cite{Lieu:2003ee}.     

In van Cittert-Zernike formalism \cite{van Cittert-Zernike}, for estimating the degree of coherence, one is integrating out the amplitude

\begin{eqnarray}\label{datsemulitalgha}\bar{\mathfrak{A}}(t,\,r) = \int\limits_{\bar{\omega}-\Delta\omega}\limits^{
\bar{\omega}+\Delta\omega}d\omega \,
 \mathfrak{A}(\omega)e^{i\left\{\left[k(\omega) -
      k(\bar{\omega})\right] \,r-\left[\omega - \bar{\omega}\right]
    t\right\}}~,\nonumber \end{eqnarray} so that the wave signal is treated as a monochromatic wave with frequency $\bar{\omega}$, wave
number $k(\bar{\omega})$ and variable amplitude $\bar{\mathfrak{A}}$,

\begin{eqnarray}\label{quasimonochromatischen}  \Phi(t,\, r) \,=\, \bar{\mathfrak{A}}(t,\,r)\,\frac{e^{i\left[k(\bar{\omega})\,r-\bar{\omega} t\right] }}{r} ~.  \end{eqnarray} With a good accuracy the source can be thought of as a plain (2D) object. Dividing the source into small elements $d\sigma_m$ and denoting by $\Phi_{m1}(t)$ and $\Phi_{m2}(t)$ the signals coming from this element at the telescope apertures, respectively, for the correlation function
between the light signals $\Phi_{1}(t)$ and $\Phi_{2}(t)$ one finds

\begin{equation} \langle\Phi_1(t)\Phi^*_2(t) \rangle = \sum\limits_m
\langle\Phi_{m1}(t)\Phi^*_{m2}(t) \rangle  ~, \nonumber \end{equation} where one takes into account that 

\begin{equation}
\langle\Phi_{m1}(t)\Phi^*_{n2}(t) \rangle =
0~,~~~~\mbox{for}~~m \neq
n ~ , \nonumber 
\end{equation} as different source elements are  mutually incoherent that means that there is
no correlation between $\Phi_{m1}(t)$ and  $\Phi_{n2}(t)$ when $m\neq n$ (the same argument was used in Eq.\eqref{Zufallsphase} as well). It is worth noticing that here $\langle \rangle$ stands for time averaging. Using an explicit expression \eqref{quasimonochromatischen}, for the correlation function one finds  

\begin{eqnarray} && \langle\Phi_{m1}(t)\Phi^*_{m2}(t)
\rangle   \,=  \nonumber \\&&  \left\langle \bar{\mathfrak{A}}_m(t,\,r_{m1})\bar{\mathfrak{A}}^*_m(t,\,r_{m2})
\right\rangle \,\frac{e^{ik(\bar{\omega})(r_{m1}-r_{m2})}}{r_{m1}\,r_{m2}}~. \nonumber \end{eqnarray} Then under the assumption  

\begin{eqnarray}\label{damkhmarepiroba} |r_{m2} \,-\, r_{m1}| \, \ll \, \left|k(\bar{\omega}\,+\,\Delta \omega) \,-\, k(\bar{\omega}) \right|^{-1} ~,  \end{eqnarray} one usually makes the following approximation \cite{Maziashvili:2012dd} \begin{eqnarray}  \bar{\mathfrak{A}}_m(t,\,r_{m1})\bar{\mathfrak{A}}^*_m(t,\,r_{m2})
 \,\approx \, \bar{\mathfrak{A}}_m(t,\,r_{m1})\bar{\mathfrak{A}}^*_m(t,\,r_{m1}) \,\approx \nonumber \\  \bar{\mathfrak{A}}_m(t,\,r_{m2})\bar{\mathfrak{A}}^*_m(t,\,r_{m2})~. \nonumber  \end{eqnarray}

Let us notice that in the standard case the auxiliary condition \eqref{damkhmarepiroba} is nothing else but the requirement the path difference to be less than the duration of the wave packet $\Delta t \simeq \Delta\omega^{-1}$. It ensures the overlapping of the waves arising from the wave packet \eqref{datsemulitalgha} after passing through the apertures. Hence, the interference effect takes place when this condition is satisfied. In the case of modified dispersion relation, which for the problem under consideration will take the form (here $\gamma$ is a numerical factor of order unity)  

\begin{equation}
\omega(k) \,=\, k \,+\, \gamma l_P^\alpha k^{1+\alpha} \,+\, \cdots ~, 
\end{equation} the velocity of the wave \eqref{datsemulitalgha} is shifted as \cite{gamma} 

\begin{equation}
v(\bar{\omega}) \,=\, \left.\frac{d\omega}{dk}\right|_{\bar{\omega}} \,=\, 1 \,+\, \gamma (l_P\bar{\omega})^\alpha \,+\, \cdots  ~, \nonumber
\end{equation} and therefore the coherence condition

\begin{equation}
\frac{|r_{m2} \,-\, r_{m1}| }{v(\bar{\omega})} \, \ll \,   \Delta\omega^{-1} ~, \nonumber
\end{equation} does no longer coincide with the Eq.\eqref{damkhmarepiroba} 

\begin{equation}
|r_{m2} \,-\, r_{m1}| \, \ll \,   \frac{1}{\Delta\omega\left|1 \,-\, \gamma l_P^\alpha\Delta\omega^\alpha\right|} ~, \nonumber
\end{equation} but the corrections are so small that they do not represent any serious interest for the problem we are discussing. Namely, the corrections are controlled by the ratio $(l_P/\bar{\lambda})^\alpha$, where $\bar{\lambda}$ is a wavelength of the light coming from high-redshift objects to the interferometer. (In a most optimistic case one can take: $\bar{\lambda}\simeq 10^{-8}$cm \cite{Ragazzoni:2003tn, Steinbring:2006ja, Tamburini:2011yf,  Steinbring:2015rla}).

\noindent So, ultimately the correlation function takes the form  
    
\begin{eqnarray} && \langle\Phi_{m1}(t)\Phi^*_{m2}(t)
\rangle   \,=  \nonumber \\&&  \left\langle \bar{\mathfrak{A}}_m(t,\,\bar{r})\bar{\mathfrak{A}}^*_m(t,\,\bar{r})
\right\rangle \,\frac{e^{ik(\bar{\omega})(r_{m1}-r_{m2})}}{r_{m1}\,r_{m2}}~, \nonumber \end{eqnarray} where $\bar{r}$ denotes the distance from the source to the telescope. The quantity $ \left\langle \bar{\mathfrak{A}}_m(t,\,\bar{r})\bar{\mathfrak{A}}^*_m(t,\,\bar{r})
\right\rangle$, which characterizes the radiation intensity from the element $d\sigma_m$, with a good accuracy may be assumed to be uniform over the source. Therefore, the corrections to the degree of coherence - characterized by the quantity \cite{van Cittert-Zernike}

\begin{eqnarray}\label{koherentulobiskhariskhi}
\left( \sum\limits_{j} \frac{\left\langle \bar{\mathfrak{A}}_j(t,\,\bar{r})\bar{\mathfrak{A}}^*_j(t,\,\bar{r}) \right\rangle}{r_{j1}^2}\right)^{-1/2} \times \nonumber \\ \left( \sum\limits_{n} \frac{\left\langle \bar{\mathfrak{A}}_n(t,\,\bar{r})\bar{\mathfrak{A}}^*_n(t,\,\bar{r}) \right\rangle}{r_{n2}^2}\right)^{-1/2} \times \nonumber \\  \sum\limits_{m} \left\langle \bar{\mathfrak{A}}_m(t,\,\bar{r})\bar{\mathfrak{A}}^*_m(t,\,\bar{r}) \right\rangle \,\frac{e^{ik(\bar{\omega})(r_{m1}-r_{m2})}}{r_{m1}\,r_{m2}} ~, 
\end{eqnarray} due to amplitude fluctuations \eqref{Neudefinition} is negligibly small.

Consider now the additional factor that comes from Eq.\eqref{Wahrscheinlichkeitsdichte}. The result given by the Eq.\eqref{Wahrscheinlichkeitsdichte} can immediately be applied to the spatial correlation function by replacing $\tau \rightarrow |\mathbf{r}_{m2}-\mathbf{r}_{m1}|$, which is of the order of the distance between the apertures. Namely, the discussion above this equation can readily be generalized to the stellar radiation by replacing the atoms with the elements $d\sigma_m$. With no loss of generality, let us assume that $r_{m2}> r_{m1}$. The wave \eqref{quasimonochromatischen}, that starts from the element $d\sigma_m$, may have an arbitrary random phase, $e^{i\theta_m(t)}$, when it reaches the point $\mathbf{r}_{m1}$. We can simplify the discussion and hold from the outset that the amplitude of this wave is constant as the radiation intensity determined by this amplitude in the expression of the coherence degree is assumed to be uniform over the whole source (see the discussion above the Eq.\eqref{koherentulobiskhariskhi}). During the time $\tau$ needed this wave to reach the point $\mathbf{r}_{m2}$ the phase may take on some other random value  $e^{i\theta_m(t+\tau)}$. So, one arrives basically at the expression \eqref{Zufallsphase}. The key observation here is that the time scale for the accumulation of effect is set by the path difference and not by the $\bar{r}$ itself as it was deemed in \cite{Lieu:2003ee}. Another important characteristic scale is $\tau_0$, see Eq.\eqref{Maßstab der Zeit}. The ratio of these two time scales looks as follows 

\begin{equation}\label{tanapardoba}
\frac{\tau}{\tau_0}   \, \simeq \,  \frac{\text{distance between the apertures}}{\bar{\lambda}}  \left(\frac{l_P}{\bar{\lambda}}\right)^{2\alpha} \ll 1 ~.  
\end{equation} So that the suppression factor is very close to $1$, see Eq.\eqref{Schwingungsverteilung}.

The effect of the modified dispersion relation $k(\bar{\omega})$ in Eq.\eqref{koherentulobiskhariskhi} is negligibly small \cite{Maziashvili:2012dd}. The above discussion already accounts for the light speed alteration because of modified dispersion relation but this question can be considered in a somewhat more generic context. Namely, one can model the problem by wave propagation in a randomly inhomogeneous medium 

\begin{equation}
\left( n(\mathbf{r})\partial_t^2 \,-\, \Delta \right) \Phi(t, \mathbf{r})  \,=\, 0 ~, 
\end{equation}  where the randomly fluctuating refractive index has the form: $n(\mathbf{r}) = 1+\varepsilon(\mathbf{r})$ \cite{Chen:2014qaa}. Let us notice that this equation bears a close similarity with that one describing the scalar field propagation in a weak gravitational field of the form  

\begin{equation}
ds^2 \,=\, \left[1+2U(\mathbf{r})\right]dt^2 \,-\, d\mathbf{r}^2 ~,   \nonumber
\end{equation} namely 

\begin{equation}\left\{\left[ 1-U(\mathbf{r}) \right]\partial_t^2 \,-\, \Delta \right\} \Phi(t, \mathbf{r})  \,=\, 0 ~.\nonumber
\end{equation} One could try to go further and study the electromagnetic wave propagation in the fluctuating background on the basis of general equation \cite{Fock}

\begin{eqnarray}
\nabla_{\mu}\left(\nabla^{\mu}\mathcal{A}^{\nu} \,-\, \nabla^{\nu}\mathcal{A}^{\mu}\right) \,=\, \nabla_{\mu}\nabla^{\mu}\mathcal{A}^{\nu} \,+\, {\tt g}^{\nu\alpha}\mathcal{R}^{\mu}_{~\delta\mu\alpha}\mathcal{A}^\delta \,=\, 0 ~, \nonumber
\end{eqnarray} which follows after using the gauge fixing $\nabla_{\mu}\mathcal{A}^\mu=0$ and the relation

\begin{equation}
\left(\nabla_{\mu}\nabla_{\nu} \,-\, \nabla_{\nu}\nabla_{\mu}\right)\mathcal{A}^\alpha \,=\, - \mathcal{R}^{\alpha}_{~\delta\mu\nu}\mathcal{A}^\delta ~, \nonumber 
\end{equation} but so far there is almost no clue motivating this study in the context of stellar interferometry \cite{Chen:2014qaa}.

\subsection{Concluding remarks}

From Eq.\eqref{tanapardoba} it follows that the effect becomes appreciable when $\tau/\tau_0 \gtrsim 1$. That, loosely speaking, might be achieved either by reducing the wave length significantly or increasing the distance between the apertures enormously. Even if one takes $\bar{\lambda}\simeq 10^{-13}$cm \cite{Steinbring:2015rla}, the distance between apertures should be $\gtrsim 100$ km the suppression factor to manifest itself. Besides, there is an importan relation that should be taken into account. Namely, for observing the diffraction pattern, in the telescope with distance $d$ between the apertures, it is required the following relation to take place \cite{van Cittert-Zernike} 

\begin{equation}
d \,\simeq \, \frac{\bar{\lambda} \bar{r}}{\rho} ~, \nonumber
\end{equation} where $\rho$ denotes the linear size of the source. Combining it with the condition $\tau/\tau_0 \gtrsim 1$, one arrives at the relation 

\begin{equation}\label{shedegi}
\bar{\lambda} \, \lesssim \, l_P \left(\frac{\bar{r}}{\rho}\right)^{1/2\alpha} ~. 
\end{equation} This relation puts a stringent bound on $\bar{\lambda}$ that can be seen immediately by taking $\alpha=1/2$ and unrealistically optimistic values $\bar{r} = H_0^{-1} \approx 10^{28}$cm (present value of the cosmological horizon) and $\rho = 10^5$cm. For those values one finds $\bar{\lambda}  \lesssim 10^{-10}$cm.

To summarize, we used a working model of the chaotic light for describing the cumulative effect, which was envisaged in \cite{Lieu:2003ee}. The basic steps of the discussion are as follows: The source is represented by the small radiating elements. The cross terms in the correlation function (which determines the degree of coherence) give a zero average contribution. The correlation function is thus determined by the sum of correlation functions for separate radiating elements. The effect of background space fluctuations is described by the overall factor which accounts for the statistical average over the randomly acquired phases for each radiating element. For estimating this factor in a semi-qualitative manner - an important role is played by the time scale at which the random phases become of the order of $\pi$. (The phase of the emitted light by an atom can be altered arbitrarily due to collisions, while the phase fluctuation of the light due to background space perturbations is tiny at a time - of the order of $(l_P\omega)^\alpha$.) If the phases acquired in a random way can take on arbitrary values, then this average factor becomes zero, see Eq.\eqref{Gleitender Mittelwert}. (For this reason, the contribution of the atoms that have experienced collisions drops out of the correlation function, see Eq.\eqref{Wahrscheinlichkeitsdichte}.) The effect for realistic parameters turns out to be negligibly small as it is plainly seen from  Eq.\eqref{shedegi}. Besides, one may address various Planck scale corrections to the coherence effect \cite{Maziashvili:2009gt, Dowker:2010pf} but from the standpoint of present (or near-future) observations they seem to be less interesting.

\acknowledgments

The author is indebted to Eric Steinbring for useful comments. This research was partially supported by the Shota Rustaveli National Science Foundation under contract number 31/89.


\begin{thebibliography}{10}
\bibitem{gamma}  
As $\gamma$ is not specified, we assume all numerical factors of order unity that pop up are absorbed in it.
\end{thebibliography}
\end{document}